\documentclass[12pt]{article}
\usepackage{amsmath, amssymb}
\usepackage{graphicx}
\usepackage{bm}

\def\keywords{\vspace{-.3em}
    \if@twocolumn
      \small\it Keywords\/\bf---$\!$%
    \else
      \begin{center}\small\bf Keywords\end{center}\quotation\small
    \fi}
\def\endkeywords{\vspace{0.6em}\par\if@twocolumn\else\endquotation\fi
    \normalsize\rm}
\def\appendix{\par
    \setcounter{section}{0}\setcounter{subsection}{0}
    \def\thesection{\Alph{section}} \section*{Appendix}
}

\newtheorem{theorem}    {Theorem}
\newtheorem{lemma}{Lemma}
\newtheorem{remark}{Remark}

\newcommand{\condset}[2]{\left\{ #1 \, \,\left|\, \, #2 \right. \right\}}

\newcommand{\defeq}{\stackrel{\rm def}{=}}
\newcommand{\bR}{\mathbb{R}}
\newcommand{\bC}{\mathbb{C}}
\newcommand{\cH}{{\cal H}}

\newcommand{\Tr}{{\rm Tr}}

\newcommand{\cHn}{{\cal H}^{\otimes n}}
\newcommand{\rhon}{\rho^{\otimes n}}
\newcommand{\sigman}{\sigma^{\otimes n}}
\newcommand{\Tn}{T_n}
\newcommand{\alphan}{\alpha_n}
\newcommand{\betan}{\beta_n}
\newcommand{\legenphi}{\Phi}
\newcommand{\legenpsi}{\Psi}
\newcommand{\qchan}{{\cal E}}
\newcommand{\cF}{{\cal F}}
\newcommand{\cG}{{\cal G}}
\newcommand{\rhovec}{\vec{\bm{\rho}}}
\newcommand{\sigmavec}{\vec{\bm{\sigma}}}

\newcommand{\bra}[1]{\langle #1 |}
\newcommand{\ket}[1]{| #1 \rangle}
\newcommand{\inprod}[2]{\langle #1 | #2\rangle}

\makeatletter
\newcommand{\lleq}{\mathrel{\mathpalette\gl@align<}}
\newcommand{\ggeq}{\mathrel{\mathpalette\gl@align>}}
\newcommand{\gl@align}[2]{
\vbox{\baselineskip\z@skip\lineskip\z@
\ialign{$\m@th#1\hfil##\hfil$\crcr#2\crcr{}_{{}_{(=)}}\crcr}}}
\makeatother

\begin{document}

\title{The Converse Part of The Theorem \\ 
for  Quantum Hoeffding Bound}

\author{
Hiroshi Nagaoka
\thanks{
Graduate School of Information Systems,
The University of Electro-Communications, 
1-5-1, Chofugaoka, Chofu-shi, Tokyo, 182-8585, Japan.
(e-mail: nagaoka@is.uec.ac.jp)
}
}

\date{}

\maketitle

\begin{abstract}
We prove the converse part of the theorem for quantum Hoeffding 
bound on the asymptotics of quantum hypothesis testing, 
essentially based on an argument developed by 
Nussbaum and Szkola in proving the converse part of 
the quantum Chernoff bound.  Our result complements 
Hayashi's proof of the direct (achievability) part of the theorem, 
so that the quantum Hoeffding bound has now been established. 
\end{abstract}

\begin{keywords}
quantum hypothesis testing,
Hoeffding bound,
error exponent
\end{keywords}

\section{Introduction}
Let $\rho$ and $\sigma$ be arbitrary density operators on 
a Hilbert space $\cH$, and consider the hypothesis testing
problem for $\rhon$ and $\sigman$.  Identifying 
a hermitian operator $0\leq \Tn\leq I$ on $\cHn$ with 
a POVM $(\Tn, I-\Tn)$ which represents 
a test of the hypotheses $\{\rhon, \sigman\}$ on 
the true state, the error probabilities of the first and second kinds 
are defined by
\[
\alphan[\Tn]\defeq 1-\Tr [\rhon \Tn]
\quad\mbox{and}\quad 
\betan[\Tn]\defeq \Tr[\sigman \Tn].
\]
Our concern is the following quantity:
\begin{equation}
B(r\,|\,\rho\,\|\,\sigma) \defeq 
\sup_{\{\Tn\}} 
\condset{
-\lim_{n\rightarrow\infty} \frac{1}{n}\log \alphan[\Tn]}
{\limsup_{n\rightarrow\infty}\frac{1}{n} \log\betan[\Tn]\leq -r
},
\label{def:B(r)}
\end{equation}
where $r$ is an arbitrary positive number. 
Since the quantum Stein's lemma established by 
\cite{Hiai-Petz} and \cite{Ogawa-Nagaoka} implies 
that 
\begin{equation}
B(r\,|\,\rho\,\|\,\sigma) = 0\quad \mbox{if}\quad r>D(\rho\,\|\,\sigma) 
\defeq \Tr[\rho (\log\rho - \log\sigma)],
\label{B(r)=0}
\end{equation}
we can assume $0<  r\leq D(\rho\,\|\,\sigma)$. 
In the classical case where probability distributions $\{p, q\}$ 
on a common discrete set $\Omega$ are 
given instead of $\{\rho, \sigma\}$, we have 
(e.g., \cite{Blahut, DemZei}), for 
$0< \forall r \leq D(p\,\|\,q)\defeq \sum_\omega p(\omega) (\log p(\omega)/q(\omega))$, 
\begin{equation}
B(r\,|\,p\,\|\,q)  = \max_{0\leq s<1}\frac{-sr-\phi(s)}{1-s},
\label{c-Hoeffding1}
\end{equation} 
where 
\[
\phi(s) = \phi(s\,|\,p\,\|\,q)\defeq \log \sum_{\omega\in\Omega} p(\omega)^{1-s} q(\omega)^s.
\]
This result is often referred to as the Hoeffding bound  after \cite{Hoe}. 
 Our aim is to show that the same expression holds for 
$B(r\,|\,\rho\,\|\,\sigma)$ as follows. 

\begin{theorem}
\label{theorem_q-Hoeffding}
For any $0<  r \leq D(\rho\,\|\,\sigma)$ we have
\begin{equation}
B(r\,|\,\rho\,\|\,\sigma)  = \max_{0\leq s<1}\frac{-sr-\phi(s)}{1-s},
\label{q-Hoeffding1}
\end{equation} 
where
\begin{equation}
 \phi(s)  = \phi(s\,|\,\rho\,\|\,\sigma)\defeq \log 
\Tr\left[ \rho^{1-s} \sigma^s\right].
\label{def_phi_quantum}
\end{equation}
\end{theorem}

Finding such a compact expression as (\ref{q-Hoeffding1}) 
for $B(r\,|\,\rho\,\|\,\sigma)$ 
has been a long standing open problem; see 
\cite{Ogawa-Hayashi} and section 3.4 of \cite{Hayashi:book} 
for significant partial results on this problem.  
Recently, two remarkable results were reported  
on the error error exponent for 
symmetric Bayesian discrimination of two quantum i.i.d.\  states, 
which had completed the theorem yielding the quantum Chernoff bound. 
That is, firstly Nussbaum and Szkola \cite{chernoff_lower} proved the 
converse part of the theorem claiming that the exponent cannot exceed the bound, 
and then Audenaert et al.\ \cite{chernoff_upper} proved the direct part 
for the achievability of the bound.  It should be noted that the 
quantum Chernoff bound is represented by the use of the same function as 
(\ref{def_phi_quantum}). 
The approach made in \cite{chernoff_upper} 
was immediately extended by Hayashi \cite{Hayashi_hoeffding} 
to the asymmetric setting, whereby he proved 
that (LHS)$\geq$ (RHS) in (\ref{q-Hoeffding1}).  In this paper 
we show the converse inequality (LHS)$\leq$ (RHS) 
based on an argument developed in \cite{chernoff_lower}. 

\section{Statement of the main result and some preliminary arguments}

Our goal is to prove that for any sequence of tests $\{\Tn\}$ 
and for any $0< r\leq D(\rho\,\|\,\sigma)$ 
the following implication holds:
\begin{equation}
\limsup_{n\rightarrow\infty} \frac{1}{n}\log \beta_n [\Tn] \leq -r 
\;\;\Longrightarrow \;\;
\liminf_{n\rightarrow\infty} \frac{1}{n}\log \alpha_n [\Tn] \geq -b(r),
\label{main1}
\end{equation}
where
\begin{equation}
b(r) = b(r\,|\,\rho\,\|\,\sigma)\defeq 
 \max_{0\leq s<1}\frac{-sr-\phi(s)}{1-s}
 \label{def:b(r)}
\end{equation}
with $\phi(s)$ defined by (\ref{def_phi_quantum}).
Let us define $\legenphi (a) $ and $\legenpsi (a) $ for 
$ - D(\rho\,\|\,\sigma) \leq a\leq D(\sigma\,\|\,\rho)$
by 
\begin{align}
\legenphi (a) &\defeq \max_{s\in\bR} \left( a s - \phi (s)\right)
= \max_{0\leq s\leq 1} \left( a s - \phi (s)\right), \\
\legenpsi (a) &\defeq \max_{s\in\bR} \left( a s - \phi (s+1)\right) = \legenphi (a) -a. 
\end{align}
Then we can see\footnote{%
For the derivation of these properties of $\legenphi$ and $\legenpsi$, 
refer to \cite{Ogawa-Nagaoka, Ogawa-Hayashi, Hayashi:book, Hayashi_hoeffding}. 
}
 that $\legenphi$ ($\legenpsi$, resp.) 
 is continuous and monotonically increasing (decreasing, resp.) on 
 the domain $[- D(\rho\,\|\,\sigma), D(\sigma\,\|\,\rho)]$ and that 
 \begin{gather*}
 \legenphi (- D(\rho\,\|\,\sigma)) = 0,  \quad 
 \legenphi (D(\sigma\,\|\,\rho)) = D(\sigma\,\|\,\rho), \\
  \legenpsi (- D(\rho\,\|\,\sigma)) = D(\rho\,\|\,\sigma), \quad 
  \legenpsi (D(\sigma\,\|\,\rho)) = 0. 
 \end{gather*}
Hence, every $r\in[ 0,  D(\rho\,\|\,\sigma)]$ is uniquely represented 
as $r = \legenpsi (a)$ by an $a\in [- D(\rho\,\|\,\sigma), D(\sigma\,\|\,\rho)]$. 
Furthermore, it can be shown that 
\begin{equation}
 r= \legenpsi (a) \;\;\Longleftrightarrow\;\; b(r) = \legenphi (a) = a+\legenpsi (a). 
 \label{r_b(r)}
\end{equation}
Therefore, the implication (\ref{main1}) for $0< r\leq D(\rho\,\|\,\sigma)$ 
is equivalent to 
\begin{equation}
 \limsup_{n\rightarrow\infty} \frac{1}{n}\log \beta_n [\Tn] \leq - \legenpsi (a)
\;\;\Longrightarrow \;\;
\liminf_{n\rightarrow\infty} \frac{1}{n}\log \alpha_n [\Tn] \geq -a -\legenpsi (a) 
\label{main2}
\end{equation}
for $-D(\rho\,\|\,\sigma) \leq a < D(\sigma\,\|\,\rho)$, 
which we shall prove in the next section. 

\section{Proof of (\ref{main2})}

Let the spectral (Shatten) decompositions of 
$\rho, \sigma$ be denoted by
\begin{equation}
 \rho = \sum_i \lambda_i\, \ket{x_i}\bra{x_i}, \quad 
\sigma = \sum_j \gamma_j\, \ket{y_j}\bra{y_j}, 
\label{spectral}
\end{equation}
and define 
\begin{equation}
p(i, j) \defeq \lambda_i\, |\inprod{x_i}{y_j}|^2, 
\quad 
q(i, j) \defeq \gamma_j\,  |\inprod{x_i}{y_j}|^2.
\label{def_p_q}
\end{equation}
Then $p$ and $q$ form 
probability distributions on the range 
$\Omega = \{(i, j)\}$ of the pair of indices $(i, j)$.  
In proving the converse part of the theorem 
of quantum Chernoff bound, Nussbaum and Szkola
 \cite{chernoff_lower} effectively used 
the following three facts on 
the relation between $\{\rho, \sigma\}$ and $\{p, q\}$. 
Firstly, we have the identity
\begin{equation}
\phi(s\,|\,\rho\,\|\,\sigma) = \phi(s\,|\, p\,\|\,q). 
\end{equation}
Secondly, the quantum i.i.d.\ extensions 
$\{\rhon, \sigman\}$ correspond to the classical 
i.i.d.\ extensions $\{p^n, q^n\}$ by 
(\ref{spectral}) and (\ref{def_p_q}). 
Thirdly, it holds for any projection $T$ that 
\begin{equation}
\alpha [T]+ \beta[T] 
\geq \frac{1}{2}\sum_{\omega\in\Omega} \min\{p(\omega), q(\omega)\}, 
\label{ineq1}
\end{equation}
where $\alpha [T]\defeq 
\Tr[\rho (I-T)] $ 
and 
$\beta[T]\defeq \Tr[\sigma T]$. 
The last one is the most ingenius finding in \cite{chernoff_lower}, 
which is derived by combining the general inequality 
\begin{equation}
 \lambda |u -v|^2 + \gamma |v|^2 \geq \frac{1}{2} |u|^2 \min\{\lambda, \gamma\}
\quad (\forall \lambda, \gamma \geq 0, \;\; \forall u, v\in\bC) 
\label{ineq2}
\end{equation}
with  
\begin{equation}
 \alpha[T] = \sum_{i,j}
\lambda_i \left|\inprod{x_i}{(I-T)y_j}\right|^2 
\quad\mbox{and}\quad 
\beta[T] = \sum_{i,j} \gamma_j \left|\inprod{x_i}{T y_j} \right|^2. 
\label{alpha_beta_rep}
\end{equation}

In the following lemma 
we present a slight extension of  (\ref{ineq3}) with 
a seemingly different form, which is more convenient for  
the present purpose. 

\begin{lemma}  For any test $0\leq T\leq I$ and 
any positive number $\delta$ we have
\begin{equation}
\alpha [T]+ \delta\, \beta[T] \geq 
\frac{1}{2} \left[
p\left\{ p\leq \delta\, q\right\} +
\delta\, q\left\{ p > \delta\, q \right\}
\right],
\label{ineq_main}
\end{equation}
where 
\begin{align*}
p\left\{ p\leq \delta\, q\right\} &\defeq 
\sum_{\omega: p(\omega) \leq \delta q(\omega)} p(\omega), \\
q\left\{ p > \delta\, q \right\} &\defeq 
\sum_{\omega:  p(\omega) > \delta q(\omega) } q(\omega).
\end{align*}
\end{lemma}

\begin{proof}
It is easy to see that (\ref{ineq2}) and (\ref{alpha_beta_rep}) 
yield 
\begin{equation}
\alpha [T]+ \delta\, \beta[T] \geq 
 \frac{1}{2}\sum_{\omega\in\Omega} \min\{ p(\omega), \delta\, q(\omega)\} 
 \label{ineq3}
\end{equation}
for any projection $T$ and any $\delta >0$.  In addition, this inequality holds for any 
test $0\leq T\leq I$, because $\min_T\left(\alpha [T]+ \delta\, \beta[T]\right) $ 
is attained by the projection, which is denoted by $\{\rho -\delta\sigma >0\}$ 
following \cite{Nagaoka-Hayashi}, onto the linear subspace spanned by 
the eigenvectors of  $\rho- \delta \sigma$ corresponding 
to positive eigenvalues \cite{Ho72, Helstrom}. 
It is obvious that (\ref{ineq3}) is equivalent to (\ref{ineq_main}). 
\end{proof}
\bigskip

Considering the $n$th i.i.d.\ case in (\ref{ineq_main}) and letting $\delta= 
e^{-nb}$ for an arbitrary $b\in\bR$, we have 
\begin{equation}
\alphan[\Tn] + e^{-nb}\betan [\Tn] \geq 
\frac{1}{2}\left[ f_n(b) + e^{-nb} g_n(b)\right] ,
\label{ineq_main_iid}
\end{equation}
where
\begin{align}
f_n(b)& \defeq p^n\{p^n \leq e^{-nb} q^n\} = p^n\left\{\frac{1}{n}\log\frac{q^n}{p^n} \geq b\right\},
\label{def:fn} \\
g_n(b)&\defeq q^n\{p^n > e^{-nb} q^n\} = q^n\left\{\frac{1}{n}\log\frac{q^n}{p^n} < b\right\}.
\label{def:gn}
\end{align}
Noting that 
\[\frac{1}{n} \log \frac{q^n(\omega^n)}{p^n(\omega^n)} 
= \frac{1}{n}\sum_{t=1}^n \log \frac{q(\omega_t)}{p(\omega_t)}
\quad\mbox{for}\quad \omega^n = (\omega_1, \ldots , \omega_n)
\]
and that $\legenphi$ is the Legendre transformation of 
$\phi (s) = E_p\left[e^{s\log q/p}\right]$, we see that 
Cram\'{e}r's theorem in large deviation theory (e.g., see 
\cite{DemZei}) yields
\begin{equation}
\lim_{n\rightarrow\infty}\frac{1}{n}\log f_n(b) = -\legenphi (b) = -b-\legenpsi (b) 
\label{cramer_f}
\end{equation}
if 
\[ b > E_p\left[\log\frac{q}{p}\right] = - D(p\,\|\,q) = - D(\rho\,\|\,\sigma). \]
Similarly, since $\legenpsi$ is the Legendre transformation of 
$\phi(s+1) = E_q\left[e^{s\log q/p}\right]$, we have
\begin{equation}
\lim_{n\rightarrow\infty}\frac{1}{n}\log g_n(b) = -\legenpsi (b)
\label{cramer_g}
\end{equation}
if 
\[ b < E_q\left[\log\frac{q}{p}\right] =  D(q\,\|\,p) = D(\sigma \,\|\,\rho). \]
Thus we obtain
\[ 
\lim_{n\rightarrow\infty}\frac{1}{n}\log \left[ f_n(b) + e^{-nb} g_n(b)\right] 
= 
-b-\legenpsi (b) 
\]
for $- D(\rho\,\|\,\sigma) <\forall b < D(\sigma \,\|\,\rho)$. 
Hence,   (\ref{ineq_main_iid}) implies
\begin{align}
-b-\legenpsi (b) 
&\leq \liminf_{n\rightarrow\infty} \frac{1}{n}\log 
\left(\alphan[\Tn] + e^{-nb}\betan [\Tn] \right) 
\nonumber \\
&\leq 
\max\left\{
\liminf_{n\rightarrow\infty}\frac{1}{n}\log \alphan[\Tn], 
\,
-b + \limsup_{n\rightarrow\infty}\frac{1}{n}\log \betan [\Tn]
\right\} .
\label{ineq:-b-legenpsi(b)}
\end{align}

Now we assume that 
\[
 \limsup_{n\rightarrow\infty} \frac{1}{n}\log \beta_n [\Tn] \leq - \legenpsi (a)
\]
for an arbitrarily fixed $a\in [- D(\rho\,\|\,\sigma), D(\sigma \,\|\,\rho))$. 
Then, 
substituting $b = a+\epsilon$ into (\ref{ineq:-b-legenpsi(b)})
with $0 < \epsilon <D(\sigma \,\|\,\rho)-a$, 
we have
\[
-a -\epsilon - \legenpsi (a+\epsilon) 
\leq 
\max\left\{
\liminf_{n\rightarrow\infty}\frac{1}{n}\log \alphan[\Tn], 
\,
-a - \epsilon - \legenpsi (a)
\right\}. 
\]
Moreover, 
since $\legenpsi$ is monotonically decreasing, 
the RHS cannot be $-a - \epsilon - \legenpsi (a)$. 
Therefore
\[
-a -\epsilon - \legenpsi (a+\epsilon) \leq 
\liminf_{n\rightarrow\infty}\frac{1}{n}\log \alphan[\Tn]. 
\]
Letting $\epsilon \downarrow 0$, we have
$ -a - \legenpsi (a) \leq 
\liminf_{n\rightarrow\infty}\frac{1}{n}\log \alphan[\Tn]$, 
which completes the proof of (\ref{main2}). 

\section{Concluding remarks}
We have proved (\ref{main2}) by extending 
an argument of Nussbaum and Szkola \cite{chernoff_lower}, 
yielding the converse part of Theorem~\ref{theorem_q-Hoeffding} 
for the quantum Hoeffding bound. 
Combined with the direct part which was proved by Hayashi 
\cite{Hayashi_hoeffding}, the theorem has been established. 
Several remarks on the theorem are now in order. 

\begin{remark}
{\em 
Besides (\ref{c-Hoeffding1}), $B(r\,|\,p\,\|\,q)$ in the classical case has
another expression: 
\begin{equation}
B(r\,|\,p\,\|\,q)  = \min_{\hat{p}: D(\hat{p}\,\|\,q) \leq r} D(\hat{p}\,\|\,p), 
\label{c-Hoeffding2}
\end{equation}
which is also called the Hoeffding bound as well as (\ref{c-Hoeffding1}). 
In the quantum case, as Hayashi showed in \cite{Hayashi:book} 
(sections 3.4 and 3.7), 
the inequality
\begin{equation} 
B(r\,|\,\rho\,\|\,\sigma)  \leq \min_{\hat{\rho}: D(\hat{\rho}\,\|\,\sigma) \leq r} D(\hat{\rho}\,\|\,\rho)
\;\;  (=: \tilde{b} (r) )
\label{q-Hoeffding2}
\end{equation}
follows from  the (strong) converse part of
the  quantum Stein's lemma \cite{Ogawa-Nagaoka}. 
The RHS can be represented as
\begin{equation}
\tilde{b} (r) = \max_{0\leq s<1}\frac{-sr-\tilde{\phi}(s)}{1-s}, 
\label{def:btilde}
\end{equation}
where 
\[ \tilde{\phi}(s) \defeq \log 
\Tr\left[e^{(1-s) \log\rho + s \log\sigma}\right] .
\]
It then follows from 
the Golden-Thompson inequality 
$\Tr\left[ e^{A+B}\right] \leq \Tr\left[ e^A e^B\right]$
that $\tilde{\phi}(s) \leq \phi(s)$, which gives 
another proof of (\ref{q-Hoeffding2}) by (\ref{q-Hoeffding1}), 
with showing that the equality in (\ref{q-Hoeffding2}) does not 
hold in general; see \cite{Ogawa-Nagaoka} for 
a similar remark on a slightly different context.  
}
\end{remark}

\begin{remark}
{\em 
Rewriting (\ref{q-Hoeffding1}) into 
\[ 
B(r\,|\,\rho\,\|\,\sigma) = \max_{t\geq 0} 
\,\bigl( - t\,r - \xi (t) \bigr) 
\]
by 
\[
\xi (t) \defeq (t+1) \, \phi\left(\frac{t}{t+1}\right) 
\]
and invoking that $\xi : [0\, \infty) \rightarrow\bR$ is a convex function with 
the derivative $\xi' (t)$ ranging over $[- D(\rho\,\|\,\sigma), 0)$, 
we can show that, for any $t\geq 0$, 
\begin{align*} 
\xi (t)  = 
\max_{0<  r \leq D(\rho\,\|\,\sigma)}
 \bigl(- t\, r - B(r\,|\,\rho\,\|\,\sigma) \bigr) 
=  \max_{r > 0}
\, \bigl(- t\, r - B(r\,|\,\rho\,\|\,\sigma) \bigr), 
\end{align*}
where the second equality follows from (\ref{B(r)=0}). 
This leads to the following conversion formula for (\ref{q-Hoeffding1}):  
\begin{equation}
\phi (s\,|\rho\,\|\,\sigma) =\max_{ r> 0} 
\,\bigl( - s\, r - (1-s)\, B(r\,|\,\rho\,\|\,\sigma)\bigr), 
\quad 0\leq\forall s\leq 1. 
\label{conversion}
\end{equation}
From the definition (\ref{def:B(r)}) of $B(r\,|\,\rho\,\|\,\sigma)$ 
it is obvious that, for any $r >0$ and any 
quantum channel (trace-preserving completely positive map)
$\qchan$, 
\begin{equation}
 B(r\,|\,\rho\,\|\,\sigma) \geq   B(r\,|\,\qchan(\rho)\,\|\,\qchan(\sigma)) .
 \label{monot_B}
 \end{equation}
 Thus (\ref{conversion}) yields 
 \begin{equation}
 \phi (s\,|\rho\,\|\,\sigma) \leq \phi (s\,|\qchan(\rho)\,\|\,\qchan(\sigma)), 
 \quad 0\leq\forall s\leq 1, 
 \end{equation}
or equivalently
 \begin{equation}
 \Tr\left[ \rho^{1-s} \sigma^s\right] \leq 
  \Tr\left[ \qchan(\rho)^{1-s} \qchan(\sigma)^s\right], 
   \quad 0\leq\forall s\leq 1. 
  \label{monot_renyi}
 \end{equation}
Two renowned proofs of this inequality are that of Uhlmann  \cite{Uhlmann} 
based on an  interpolation theory and that of Petz \cite{Petz:quasi} 
which, in this case,  relies upon  the operator concavity of the function 
$f(u) = u^s$.  Our proof seems to be new.  
Note that the monotonicity of quantum relative entropy 
\begin{equation} 
D(\rho\,\|\,\sigma) \geq D(\qchan(\rho)\,\|\,\qchan(\sigma))
\end{equation}
is obtained by differentiating (\ref{monot_renyi}) at 
$s=0$. 
}
\end{remark}

\begin{remark}
{\em 
For an arbitrary $a\in \bR$, let 
\begin{align}
 F_n (a) &\defeq \Tr\left[
\rhon \left\{ \rhon - e^{-na}\sigman \leq 0\right\}\right], \\
G_n (a) &\defeq \Tr\left[
\sigman \left\{ \rhon - e^{-na}\sigman > 0\right\}\right], 
\end{align}
where $\{A\leq 0\}$ ($\{A > 0\}$, resp.) for a hermitian operator $A$ 
is defined as the projection onto the subspace spanned by 
the eigenvectors of $A$ corresponding to nonpositive (positive, resp.) 
eigenvalues.   
Assume that the following limits exist:
\begin{align}
\cF (a) &\defeq - \lim_{n\rightarrow\infty} \frac{1}{n}\log F_n(a), \\
\cG (a) &\defeq - \lim_{n\rightarrow\infty} \frac{1}{n}\log G_n(a), 
\end{align}
Then, as is shown in \cite{Nagaoka-Hayashi}\footnote{%
Note that our $B(r\,|\rho\,\|\,\sigma)$ 
corresponds to  $B_e (r\,|\,\sigmavec\,\|\,\rhovec )$ in 
\cite{Nagaoka-Hayashi}. }
 we have
\begin{align*}
B(r\,|\,\rho\,\|\,\sigma) = \sup_{a: \cG (a)\geq r} \cF (a) 
= 
\inf_{a:\cG (a) < r } \left( a + \cG(a) \right). 
\end{align*}
On the other hand, since $\legenphi$ ($\legenpsi$, resp.) is continuous 
and monotonically increasing (decreasing, resp.), 
it follows from (\ref{q-Hoeffding1}) and (\ref{r_b(r)}) that 
\[
B(r\,|\,\rho\,\|\,\sigma) = \sup_{a: \legenpsi (a)\geq r} \legenphi (a) 
= 
\inf_{a:\legenpsi (a) < r } \left( a + \legenpsi (a) \right). 
\]
Comparing these expressions, we are led to the following conjecture:
\begin{equation}
\cF (a) = \legenphi (a) \quad\mbox{and}\quad 
\cG (a) = \legenpsi (a) .
\label{conjecture}
\end{equation}
If $\rho$ and $\sigma$ commute, 
these relations are equivalent to (\ref{cramer_f}) and (\ref{cramer_g}), 
and hence are true.  In the general case, however, 
they have no mathematical proof at present. 
}
\end{remark}

\end{document}